\documentclass[nologo,url,11pt,a4paper]{ETHpaper}
\usepackage{amssymb,amsmath} 
\usepackage[tight]{subfigure}
\usepackage[sort&compress]{natbib} 
\usepackage[utf8]{inputenc}
\usepackage{graphicx}
\usepackage{subfigure}
\newcommand{\rud}{R\&D }

\newcommand{\knd}{$\hat{K}_n(d)$ }
\usepackage[font={small}]{caption}
\begin{document}
\title{The spatial component of R\&D networks}
\titlealternative{The spatial component of R\&D networks}
\author{Tobias Scholl$^{\dag,\ddag}$, Antonios Garas$^\S$ and Frank Schweitzer$^\S$}
\authoralternative{Tobias Scholl, Antonios Garas  and Frank Schweitzer}
\address{
$^\dag$Philipps University of Marburg, Department of Geography\\
$^\ddag$House of Logistics \& Mobility (HOLM) GmbH, Frankfurt \\
$^\S$Chair of Systems Design, ETH Zurich, Weinbergstrasse 58, 8092
Zurich, Switzerland}
\reference{Working Paper: version of \today }
\makeframing
\maketitle
\begin{abstract} 
We study the role of geography in \rud networks by means of a quantitative, micro-geographic approach.
Using  a large database that covers international \rud collaborations from 1984 to 2009, we localize each actor precisely in space through its latitude and longitude.
This allows us to analyze the \rud network at all geographic scales simultaneously. 
Our empirical results show that despite the high importance of the city level,  transnational \rud collaborations at large distances are much more frequent than expected from similar networks.
This provides evidence  for the ambiguity of distance in economic cooperation which is also suggested by the existing literature.
In addition we test whether  the hypothesis of  \emph{local buzz and global pipelines} applies to the observed \rud network by calculating well-defined metrics from network theory.
\end{abstract}
\section{Introduction}
\label{sec:intro} 
Networks and their evolution over time have been studied extensively from different academic perspectives.
Recently, much effort has been devoted to understand the structure of economic networks from a complex systems view.
The complex systems approach, rooting back to physics and computer science, provides new perspectives on economic questions, since it allows to observe both the micro level (behavior and properties of a single economic actor) and the macro level (global outcomes, dynamics) in a unified way~\citep{Schweitzer24072009}.
Different economic networks have been the topic of recent studies, such as ownership, financial, trade, or collaboration networks \citep{Garas2008b,Garas2010,Glattfelder2011,Battiston2012,Tomasello2013}. 
Whereas this view led to the discovery of some stylized facts about economic networks, the actual role of geography in their formation  requires further investigation. 

This is quite surprising since geography has been a major topic of research in both economics and network science. 
From the economic perspective, it is widely agreed upon that geography has a prominent role in the formation of economic networks. 
Spatial proximity eases face to face contact, reduces costs and favors the building of trust. 
Geographers, economists and sociologists have provided a rich body of literature on how spatial proximity influences economic structures, starting from Marshall's Industrial Districts \cite[]{marshall1895principles} to Porter's Cluster concept \cite[]{Porter98} and Urban Economics \citep{Glaeser1992, glaeser1999learning}, just to mention a few. 

While all these theories stress the importance of spatial proximity, there is also a bulk of literature that favors the opposite: 
Internationalization, standardization, global transport routes and new communication technologies constantly reduce the costs for transnational cooperation. 
At least since the end of 1980s, there is a strong trend towards international \rud collaborations~\citep{Kang, Gassmann, Georghiou1998611}.
In economic theory, the ambiguous role of geographic distance has been labeled with the term \textit{Glocalization}, describing the high relevance of a firm's local environment in a globalized economy. 
A third stream  of literature that can be labeled as ``regional innovation system studies'' takes an intermediate point of view; 
it puts special emphasis on the 'region', normally larger than a city and smaller than a nation, as the driver of innovation, interaction and learning
\citep{cooke2001regional,doloreux2004regional}.
 
Despite these theoretical insights, the existing literature -- as far as we know -- does not provide any empirical analysis of \rud collaborations at \emph{all spatial scales, simultaneously}.
Existing studies focus on one specific geographical level, such as cities \cite[]{Acs}, regions~\citep{Fritsch2004245, EBode} or nations \cite[]{Coe12}. 
However, this view is incomplete since the nature of \rud networks allows the co-existence of collaboration at different distances.

Additionally, it is considered a ``stylized fact''in economics that \rud activities depend both on the type of industries and on properties of the firms, i.e. their size or their growth rates \citep{Klepper,deJong}, and this is reflected in the spatial component of their collaborations.
For instance, small and medium-sized firms are usually more dependent on spatial proximity within their knowledge-acquisition \cite[]{ Davenport2005683}.
Concerning the type of industry,  literature has paid much attention on the role of service firms, in particular Knowledge-Intensive Business Services (KIBS), as catalysts of innovation \cite[]{Strambach}.
For KIBS, \rud collaborations are characterized by cyclical interactions and a high importance of tacit knowledge that favors the establishment local collaborations \cite[68-70]{Muller200964}. 
Besides the firm and industry specific properties, the temporal component should have a notable influence as well.
As  literature suggests, we expect to find a shift towards international \rud collaborations with our analysis, given that we have \rud collaboration data spanning a time period of 25 years. 

In sum, the spatial component of \rud collaborations seems to be rather heterogeneous and multifaceted.
Whereas the existing literature lacks empirical investigations, this issue has been tackled from a theoretical perspective in the paper of \cite{Bathelt}.
By means of their hypothesis of \emph{local buzz and global pipelines}, the authors explain the \emph{co-existence} of different spatial scales in collaboration networks. However, as far as we know, the existing literature does not provide an empirical analysis of this theory at a global scale.

With this paper we address the role of geographic distance in \rud collaborations by providing a detailed analysis of the spatial component of \rud networks. 
Our contribution to the current literature is twofold.
First, we analyze the network of \rud alliances, which, in order to function properly, requires ongoing coordination and trust \cite[]{Kang}. 
These are complex social processes that should have traceable effects in the network's spatial component.
Second, we push the geographic scale to a micro level, using refined localization for each firm in our dataset.
This treatment allows to investigate the spatial component of \rud networks for all different spatial scales simultaneously, and goes beyond what has been reported so far in  literature. 
In addition, we are able to use well-defined network measures (like betweenness centrality) to test the existence and the function of the \emph{local buzz and global pipelines}, which are expected to play an important role in the process of knowledge creation~\cite[]{Bathelt}.
In sum, we contribute to the literature on \rud networks by focusing on the quantitative evaluation of one specific aspect: the geographic distance between collaborating firms.

\section{Literature}
\label{sec:role}
Studies of the spatial component of networks are actually not new, but have been a research topic of many early quantitative geographers in the 1960s \citep{ haggett1967models, haggett1969network}.
However, they were limited due to lack of access to large databases and computational power \cite[3]{Barthelemy}. 
Today, user- and commercially generated databases provide new possibilities to investigate the spatial component of complex networks, shedding light to the way geography affects human behaviour.

During the last two decades, regional scientists have provided a remarkable amount of literature that applies network theory in the context of spatial innovation studies.
Starting from the insight that geography and the way  agents are connected to each other have a remarkable impact on innovation activities of firms or regions  \citep[e.g.][]{acs1992real,feldman1994geography,Porter98}, scientists have included network theory in their investigations to model innovation networks in a more reliable way.
Much of this work can be labeled with the term \emph{regional regression-based network theory}: 
network measures are computed and then integrated in a regression model to determine the factors why one region is more innovative than others \citep[e.g.][]{broekel2014modeling, graf2011performance, paier2011determinants, whittington2009networks}.

While our paper also deals with networks, space and innovativeness, there are two main differences to the aforementioned stream of literature.
First, our operational unit is not the ``region'' but a single actor, i.e., a node in a network representation, and its geographical position is given by its latitude and longitude.
Second, our empirical method does not rely on a spatial econometric approach where network measures are included in regression models. 
Instead, we focus our analysis on a very basic but nonetheless often disregarded question: What is the geographic distance of nodes inside a collaboration network and which conclusions on industrial innovation mechanisms can be derived from such an analysis?
More precisely, we study the distribution of distances  between nodes in the observed \rud network.
In contrast to the mentioned stream of literature, such an analysis is more common in physics and complex systems literature.
For example, \cite{Brockmann} 
analyzed the circulation of dollar bills, in order to investigate human travel for all different modes of transport simultaneously. 
\cite{Gonzalez2008} analyzed data from mobile phone users to explore patterns in human mobility, while \cite{lambiotte2008geographical} used mobile phone data to study  the geographical component of human communication. 
Similarly, \cite{bianconi2009assessing} studied the role of distance for the global airline network, \cite{Battiston2011} studied how distance affects corporate ownership networks, and
\cite{Hennemann} collected geocoded information from the Thompson Scientific\textregistered \ database to explore the influence of geography on scientific collaborations.

All these works found that distances between nodes in the network follow a heavy-tailed distribution. 
More precisely, in most of them the distribution of distances $r$ decays as a power law 
\begin{equation}
P(r) \sim \frac{1}{r^{\gamma}}
\end{equation}
with some small variations in the exponent, $\gamma$.
As extensively discussed in literature, such broad distance distributions have important implications. 
For example, \cite{Hennemann} showed that national borders still have the highest impact on where scientific collaboration takes place. 
This contrasts with the popular belief that scientific collaboration has evolved into a joint global process and, as such, becomes independent from spatial distances.
But, while distance plays a significant role at the national level,  it is almost irrelevant for cross-country collaboration at the same time.
Therefore, the authors conclude that low coordination costs and national funding systems tie researchers to local science clusters. 
But, going beyond the national level, increasing costs seem to be unimportant when collaborators can benefit from access to complementary knowledge or equipment \cite[223 f.]{Hennemann}. 

Due to the functional similarities between scientific collaborations and \rud activities, heavy-tailed distributions of distances seem to be plausible for \rud networks as well.
This means that most of the collaborations take place between partners that are geographically close to each other, because collaborations usually imply that the economic actors know each other and meet frequently.
However, collaborations are not fixed only to a narrow geographical scale. 
Large distances between actors may be more cost-intensive but also increase the chance of obtaining complementary knowledge. 

Despite this intuitive explanation, economic theory provides only few models to explain a heavy-tailed distribution of distances in \rud collaborations, since most papers only deal with  specific geographic levels. 
\cite{Bathelt} criticize the sole focus of many theories on localized knowledge spillover, since geographical proximity alone cannot explain the information flows within clusters of firms.
Notwithstanding the importance of the surrounding environment, even very successful clusters, such as Silicon Valley, depend on knowledge flows from other remote clusters.
The authors explain this ambiguity by means of what they call {\it local buzz and global pipelines}.
Although their paper deals with the conversion of explicit and tacit knowledge and is not about network theory from quantitative point of view, it is often cited in the stream of regional regression-based network theory \citep[e.g.][]{graf2011performance, morrison2009knowledge, whittington2009networks}.

\begin{figure} 
\centering
\includegraphics[width=.7\textwidth]{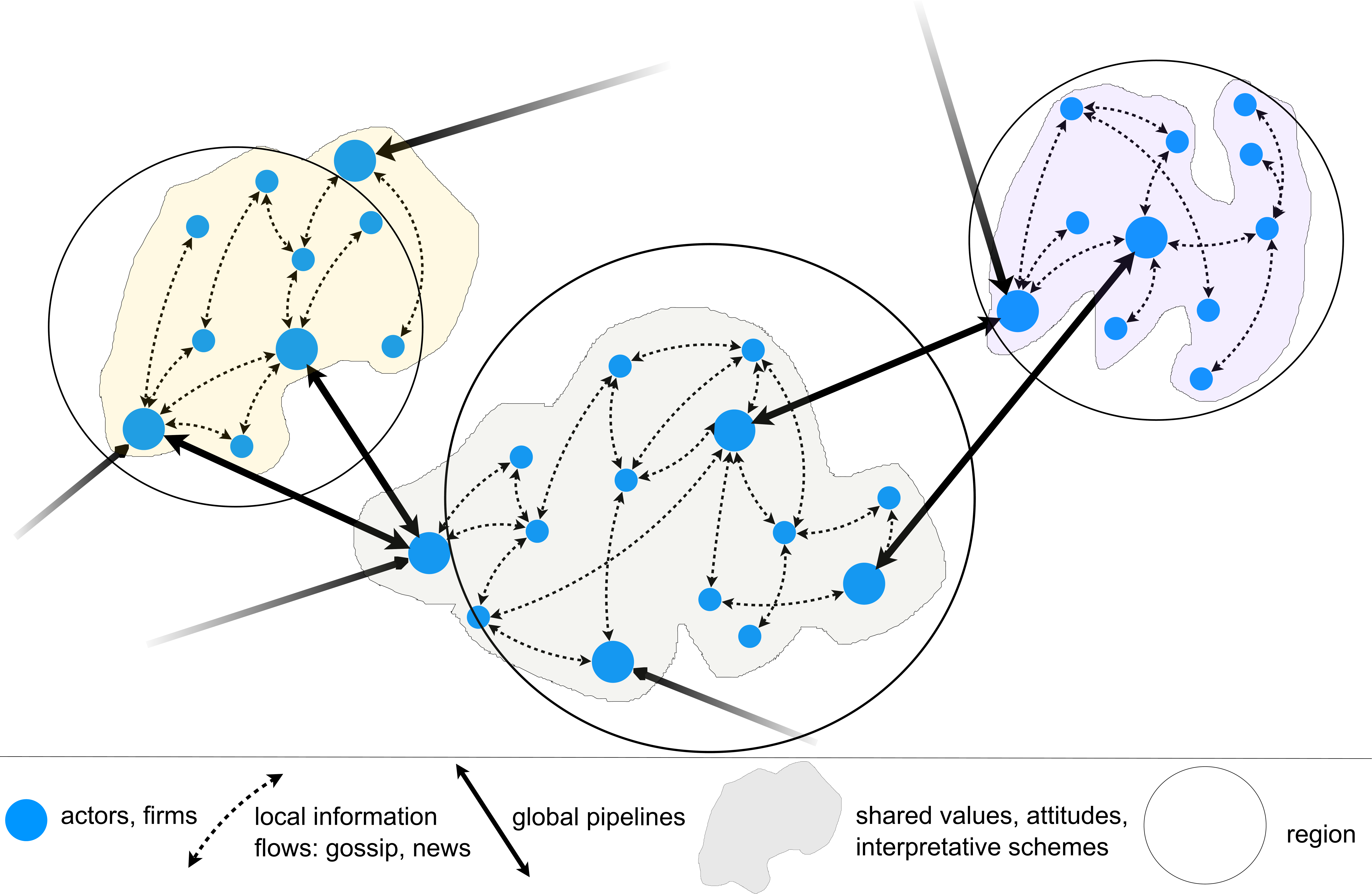}
\caption{
Graphical representation of the local buzz and global pipelines model, following \cite{Bathelt}.}
\label{fig:localbuzz}
\end{figure}

The term \textit{local buzz} stands for short distance flows of information, face-to-face contact, gossip and every day communication of people and firms within the same region.
Local buzz can be seen as a relatively informal way of communication that does not require particular investments and that affects all firms within a local cluster.
This buzz can lead to common locally shared routines, similar technology attitudes and trust \cite[38]{Bathelt}. 
On the other hand, \textit{global pipelines} represent connections of individual firms inside a local cluster to other places around the world, as illustrated in figure~\ref{fig:localbuzz}.

In contrast to the local buzz, global pipelines represent a more formal way of interaction and can be seen as strategic partnerships. 
They are highly relevant for local clusters, since a cluster's competitive advantage is not only determined by its own innovative capacity but also by its ability to quickly absorb and adapt technology created elsewhere \citep[43 ff.]{Bathelt}. 
Local buzz and global pipelines should not be seen as two distinct ways of accessing knowledge flows, but as aspects interacting with each other. Knowledge gained through global pipelines is accessible to all firms in a local cluster through the local buzz \citep[46]{Bathelt}.
In the paper of \cite{Bathelt} it is not clear, whether two collaborating firms of a global pipeline are both located inside a local cluster.
However the related literature clearly favours the idea of global pipelines being bridges between remote clusters~\citep{Amin, Giuliani}, since a costly pipeline is more efficient when it can benefit from local buzz from both of its sides.  

The two aspects of the local buzz and global pipelines model seem to provide a promising theoretical explanation for observing heavy-tailed distributed distances of \rud collaborations:
Local buzz is easy to access and facilitates the development of local trust, which is highly important for \rud collaborations \cite[]{RYAN}. 
Thus, local buzz can explain the strong focus of \rud collaborations on short distances.
In order to adopt crucial knowledge created elsewhere, firms also invest in building up global pipelines. 
However, global pipelines are less frequent since they require intense efforts on both sides, are costly and harder to establish and to maintain.

Because the \emph{local buzz and global pipelines} hypothesis describes the heterogeneity of distances inside a network in a convincing manner, we plan to test it against our dataset and derive conclusions about the way \rud networks are structured.
However, observing a broad distribution of distances  alone does not imply that a \rud network follows a \emph{local buzz and global pipelines} pattern.
It only shows that remote connections exist but we do not know whether they bridge local clusters.
If a network is characterized by separated local clusters with some pairs of firms being bridges that bring together different clusters, as shown in Figure~\ref{fig:localbuzz}, it should also have a heavy-tailed distribution of the network's edge betweenness centrality ($b_c(e)$) values (see section \ref{sec:methods}). 
But, this will only show the existence of pipelines which by construction should have high $b_c(e)$ values. To actually test how global they are, we should be able to detect correlations between their $b_c(e)$  values and their length in terms of actual geographic distance.

\section{Data}
For our analysis, we use the Thomson Reuters SDC Platinum database.
This database records all publicly announced \rud partnerships between different kind of economic actors (including firms, investors, banks and universities) for the time period of 1984 to 2009.
Hereafter, we use the standard view of complex networks literature, to represent economic actors as nodes and any relations between them as edges in an \rud network.
In general, these edges can be directed or undirected. 
Given that \rud collaborations are usually bilateral processes, we assume that the network is undirected.
With this abstract view we map actor differences to node properties, while all aspects of their relationship (like number of alliances between the same set of nodes, their geographical distance, etc.) are mapped to edge properties.

For each alliance relation we know the date of its announcement, the participating nodes, their 4-digit SIC~\footnote{Standard Industrial Classification.} codes, and their country of residence.
Because the nodes' geographic information is highly aggregated in our database, we used on-line queries in search engines to retrieve the address (house number, street name, postal code and city) of 3598 nodes.
This allows us to calculate their exact spatial position through the geographical latitude and longitude. 
This way, we circumvent the Modifiable Areal Unit Problem (MAUP), a well-known issue in spatial statistics that describes a general bias when working with spatially aggregated data~\cite[]{Openshaw}. 
For our type of analysis, even small scaled spatial units may lead to distortion when processing global data. 
For instance, postal codes that are quite often used for geographically detailed research, normally  refer to a single dwelling in the UK, while they range between 12 and 1,400 square kilometers in the US~\cite[]{Grubesic}. 
In summary, knowing the exact location of each node enables us to conduct a global analysis for all spatial scales, simultaneously.
Additional to the 4-digit SIC-code, we matched our nodes with the Bureau Van Dijk’s 2007 Orbis database to obtain information about its size, as reflected by the number of employees.

\section{Methods}
\label{sec:methods}
\paragraph{Centrality measures}
A graph $G$, is an abstract representation of what in this work we call network, and it is defined as a tuple of vertices (or nodes) $n\in V$ and edges $e\in E$ connecting the nodes.
The way nodes are connected, i.e. the topology of the network, determines its properties, which can be expressed by  various network metrics.  The most frequently used metrics are related to \emph{centrality}~\cite[]{newman2010}, which can be defined in  many different ways. In this work we will only consider  node degree centrality and  edge betweenness centrality.
The degree centrality of a node is simply given by its number of edges, $k$, and is called node degree.
The edge betweenness centrality is given by
\begin{equation}
b_c(e) = \sum_{s \neq t} \frac{\sigma_{st}(e)}{\sigma_{st}}
\end{equation}
where $\sigma_{st}$ is the total number of shortest paths from node $s$ to node $t$, and $\sigma_{st}(e)$ is the number of those paths passing through edge $e$.
This centrality metric is of particular interest to identify edges that act as pipelines, since by definition edges with high betweenness centrality are those that act as bridges between different communities. 

\paragraph{Null model}
For different purposes, we have to compare the outcome of the observed \rud network to that of a random network.
For instance, in order to make statements on the pipelines properties, we have to compare the distribution of $b_c(e)$  values of the observed \rud network to a random one, keeping  the main properties of the network (number of edges and nodes, degree distribution)  constant except for the allocation of edges to nodes.  
To construct such a network, we randomly pick two edges, $e_{ij}=(n_i,n_j)$ and $e_{kl}=(n_k,n_l)$, and swap the connections to obtain $e_{ik}=(n_i,n_k)$ and $e_{jl}=(n_j,n_l)$.
This step  has to be repeated at least as many times as there are edges in the network. 
Here, following \cite{Hennemann}, we repeat this procedure five times the observed number of edges to ensure a sufficient randomization.

\paragraph{Point pattern analysis}
Since our dataset provides refined information about the location of firms, it
can be studied with techniques that have been established for point pattern analysis. 
For the point pattern analysis of economic activities, the $\hat{K}(d)$-index of \cite{Duranton} is the most established one in spatial econometrics. 
$\hat{K}(d)$ allows to test whether, and at which exact distances, firms in an industry under investigation are spatially more concentrated or dispersed relative to what would be expected from the overall industrial agglomeration. 
The first step to compute  $\hat{K}(d)$-values is to build the geographical distances between all possible pairs of firms so that one gains $N(N-1)/2$ unique bilateral distances, where  $N$  is the number of firms in the observed industry. 
In the next step, one estimates the density of neighborhoods of firm pairs at distance $d$ by means of a kernel density estimation (KDE).
Thus the formula reads:
\begin{equation}
\hat{K}(d) = \frac{1}{n*(n-1)h}\sum_{i=1}^{n-1} \sum_{j=i+1}^{n}f\left(\frac{r-d_{i,j}}{h}\right),
\end{equation}
where $d_{i,j}$ is the distance between firm $i$ and $j$, $h$ is the optimal bandwidth and $f$ is the Gaussian kernel function.
To test whether the computed  $\hat{K}(d)$ values show significant concentration or dispersion at a distance $d$, they have to be compared to confidence bands that are constructed by a Monte-Carlo approach: 
Let $N$ be the number of firms in the industry under investigation, then we draw $N$ locations out of the population of all possible firm locations in the area under investigation. 
These firms represent a random industry localization, whose  $\hat{K}(d)$ values  are computed.
The basic idea behind this procedure is that the spatial localization of industries does not follow a pure random pattern, since industries cannot settle anywhere in a country. 
It is obvious that natural barriers (lakes, rivers, mountains) or political restriction (nature reserves, residential areas) limit the location choice of entrepreneurs \cite[1085]{Duranton}. 
The procedure of drawing random firm locations is repeated 1000 times, storing the  $\hat{K}(d)$ values of each iteration. 
For each estimated distance interval $d$, the 5$^{th}$ and the 95$^{th}$ percentile are selected and are used to provide us with the lower (5$^{th}$ percentile) and a upper (95$^{th}$ percentile) confidence band.
Whenever the $\hat{K}(d)$ value of an observed point pattern hits the upper/lower confidence band, the pattern has significantly more/less neighborhoods at the respective distance than expected from a random distribution of firm locations. 

In our case, $\hat{K}(d)$ is a useful tool to check whether \rud collaborations are significantly  concentrated at short distances, as supposed by the concept of \emph{local buzz}.
However, $\hat{K}(d)$ computes distances between all possible pairs of firms and therefore neglects the network topology.
Thus we rewrite it as $\hat{K}_n(d)$ and consider distances between firm $i$ and $j$ only if an edge $e(ij)$ exists.
Thus the formula reads:
\begin{equation}
\hat{K}_n(d) = \frac{1}{Eh}\sum_{i=1}^n \sum_{j, e(ij)\in E }^{n}f\left(\frac{d-d_{i,j}}{h}\right)
\end{equation}
where $E$ is the number of edges.

Besides, also the calculation of confidence bands has to be adopted to the case of a network:
By simply selecting random  firm locations worldwide, results might be skewed by the fact that the locations of firms involved in \rud activities are actually highly restricted  (see Figure~\ref{world}).
By constructing random networks using the null model as described above, we keep all main properties of the network (number of edges and nodes, degree distribution) unchanged, except for the length of the edges. 
Furthermore, we ensure that the locations of the nodes are that of firms with an \rud activity.   
Analogous to the procedure of  calculating $\hat{K}(d)$, we generate 1000 random networks and store the $\hat{K}_n(d)$ values of each iteration. This provides us with the lower (5$^{th}$ percentile) and a upper (95$^{th}$ percentile) confidence band.
Whenever the $\hat{K}_n(d)$ curve of the  observed \rud network hits the upper/lower confidence band, the network has significantly more/less edges at the respective distance than expected from a random distribution of edges at a 5 \% confidence level. 

\section{Empirical Analysis}
\label{sec:results}
\begin{figure} 
\centering
\subfigure[]{
\includegraphics[width=.53\textwidth]{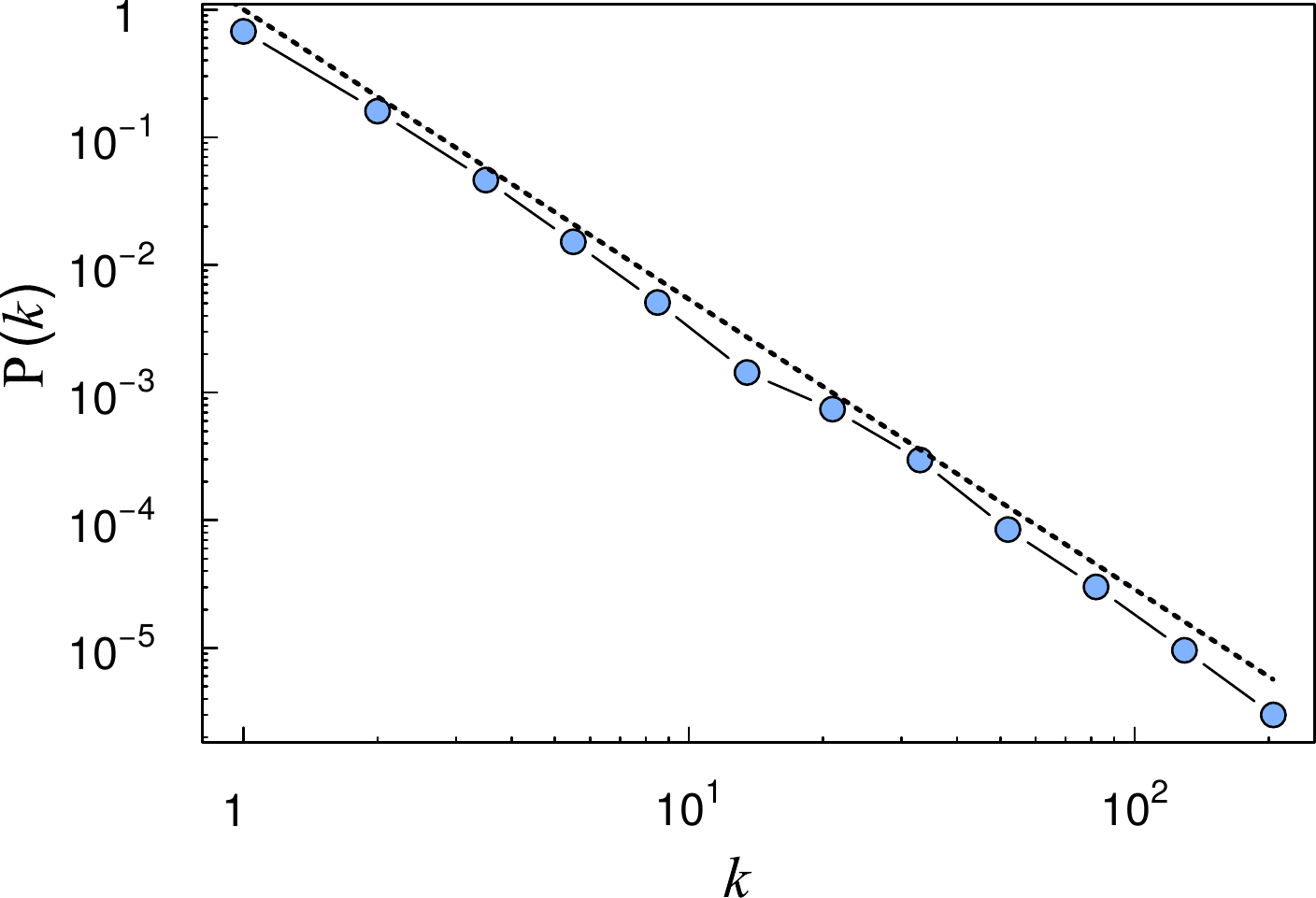}
\label{world}
}
\hfill
\subfigure[]{
\includegraphics[width=.42\textwidth]{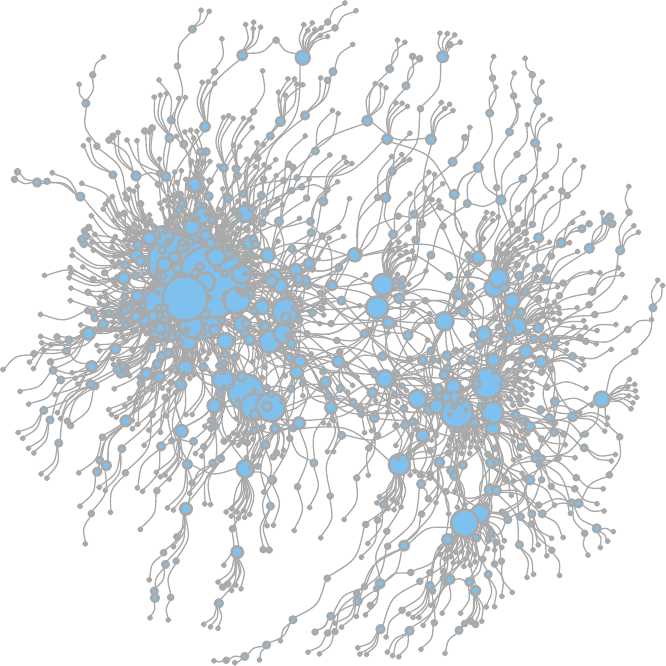}
}
\caption{Topology of the observed \rud network. Left: The degree distribution; the dashed line shows a plot of a power law function with exponent $\gamma$ = $2.27$, i.e. equal to the exponent that describes the real network. 
The null hypothesis of the KS-test (both samples were drawn from the same distribution) can not be rejected ($p$=0.99999).
Right: The largest connected component of the \rud network in an abstract space using the Fruchterman-Reingold Algorithm. The size of the nodes is relative to their degree $k$.}
\label{topo-plots}
\end{figure} 

We start with the description of the topological features of the network.
As shown in Figure~\ref{topo-plots}, the degree distribution of the \rud network follows a power law with exponent $\gamma=2.27$, obtained using the maximum likelihood method as described in ~\cite{Clauset}.
Such degree distributions characterize a class of networks that are called \textit{scale-free}, which includes biological networks, social networks, economic networks, technological networks, etc.
A large empirical analysis of the dataset used here has already been performed by \cite{Tomasello2013}.
This analysis showed the existence of broad degree distributions together with other network properties like core-periphery structure, nested organization, and small world topology. 
We should note that in most cases such properties depend on each-other, and the existence of one usually implies the existence of the others.
For example, broad degree distributions imply small network diameter which is associated with small world properties (see~\cite{cohen2003scale}), and these properties play an important role for innovation networks and knowledge flows~\citep{boschma2010handbook}.
In their analysis \cite{Tomasello2013} showed that most network properties are invariant across pooled and sectoral networks, but they do change their value across time following a non-monotonic trend.
However, in this empirical work the authors did not consider the role of geography, which is the topic of the current analysis.

\begin{figure} 
\centering
\includegraphics[width=\textwidth]{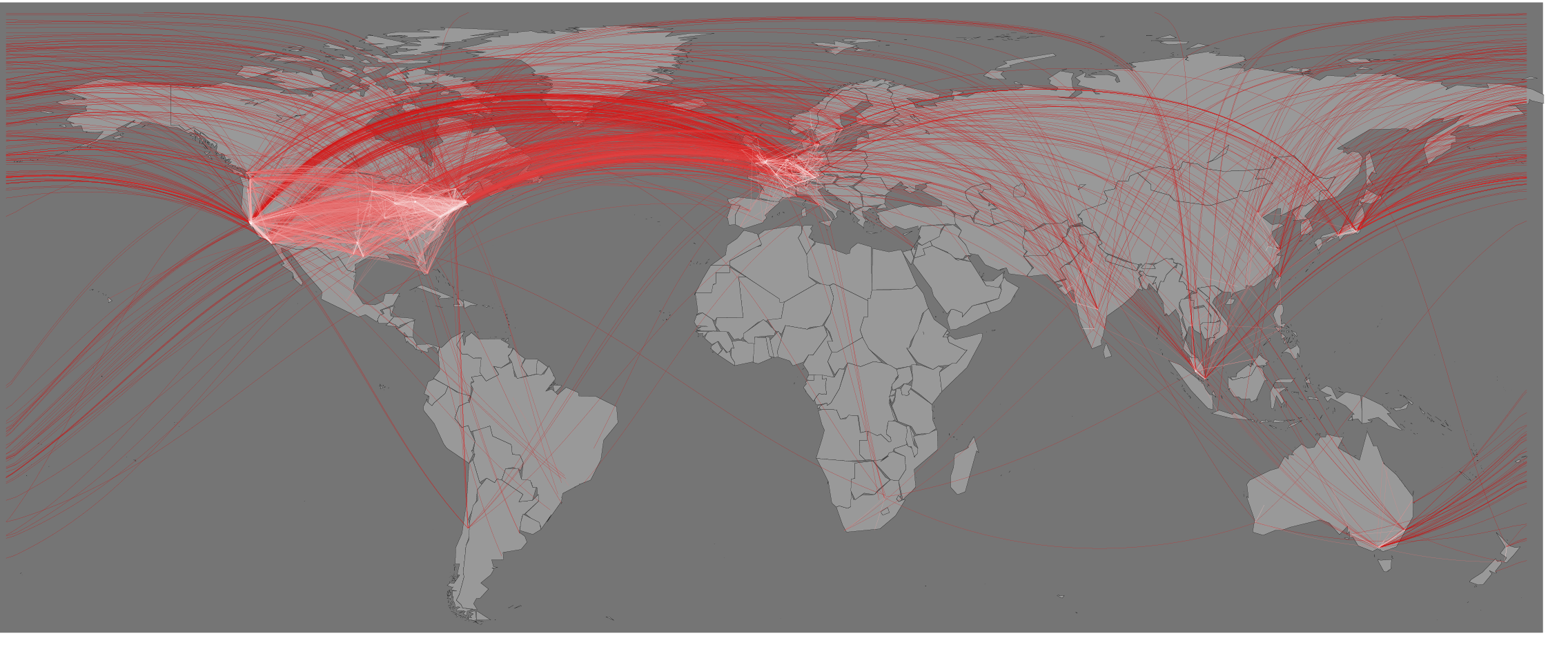}
\caption{
The global \rud network from 1984 to 2009. Nodes represent firms, edges represent \rud alliances. Edges are colored according to the orthodromic distance of their respective nodes.
}
\label{world}
\end{figure} 

Thus, hereafter we focus on the spatial dimension of the \rud network, i.e. the location of nodes and the edge lengths.
Figure~\ref{world} plots the nodes and edges of all observed collaborations  from 1984 to 2009. 
We observe a strong presence of edges between nodes (firms) located inside the US and between firms located in the US and the EU.
There are less edges to firms in Asia (Japan, India, China),  while in huge parts of the world (mainly in Southern America, Russia and Africa) only a few or not even one single node of the \rud network are located.
Focusing on short distance collaborations only (lighter colored edges), we verify that the locations of their corresponding nodes
are even more restricted and exclusive. 
There is strong concentration of such nodes at the US East and West-Coast, at the great lakes area, while in Europe high concentration is observed in Great Britain, France, and Germany. 
In sum, the described results are an expectable outcome and they favor the idea of  a broad distribution of distances. Furthermore the map suggests the existence of \emph{local buzz and global pipelines}, since we observe several global hot spots with nodes that show both edges at a local level and edges at a global level, connecting them to other global hot spots.  

\begin{figure}[t] 
\centering
\includegraphics[width=.47\textwidth]{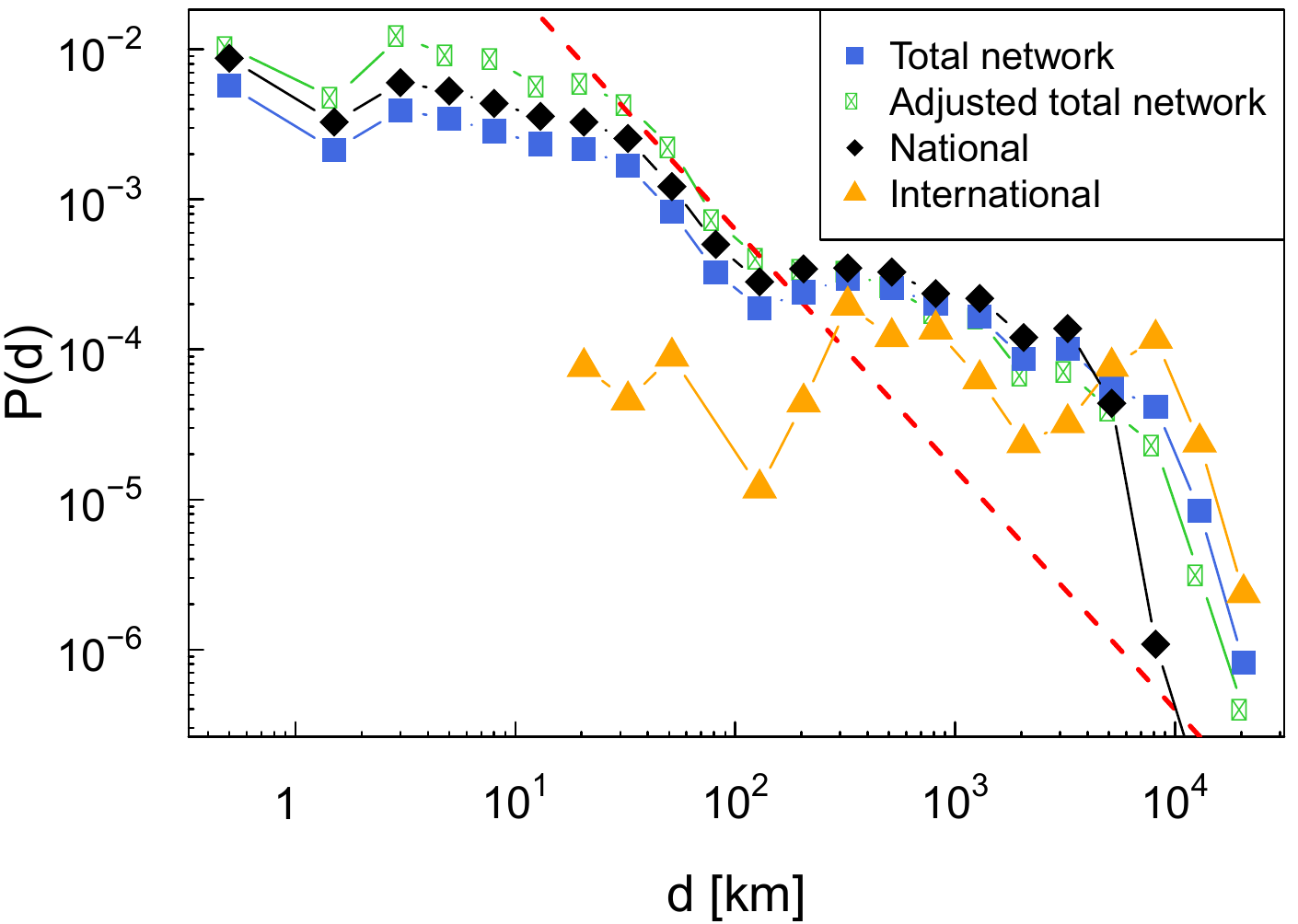}
\caption{
Distribution of edge lengths for the total network (blue) the adjusted total network (green), national cooperation (black) and international cooperation (orange). The dashed red line shows a power law distribution with $\gamma = 1.6$, as was reported by \cite{Brockmann}.
}
\label{pdf-plots}
\end{figure}

However, as shown in Figure~\ref{pdf-plots}, the distribution of edge lengths in the \rud network 
is significantly different from the distribution of edge lengths in scientific cooperation \cite[]{Hennemann} or human mobility \cite[]{Brockmann}. 
The power-law hypothesis has to be rejected ($\rm{p}<0.001$) since the observed \rud network has a much higher probability of the appearance of edges at larger distances.
This outcome can be explained by the strong presence of transnational collaborations in the observed \rud network.
Since we know the exact address of each node, we can easily compute the share of edges where both nodes originate from different countries (35 \%).
In comparison to scientific collaborations where the  level of national collaborations exceeds the international level by a factor of 10-50 \cite[217]{Hennemann}, 
\rud activities are thus clearly more international. 
When plotting the distribution of edge lengths of national and international collaborations separately, we observe a similar outcome as described by \cite{Hennemann}:
Distance has only an influence at the national level but is irrelevant for international \rud activities.

Since power-law distributed distances have been reported for many networks, we try to explain why this is not the case for the observed \rud network.
The most intuitive explanation is that this outcome reflects the nature of \rud collaborations in which the   costs of long-distance collaboration are negligible compared to the importance of accessing complementary knowledge.
However, we also have to bear in mind that the data collection of the SDC platinum database might be biased towards \rud activities of large US/EU firms.
Since the SDC database contains only publicly announced \rud alliances, it is possible that many \rud activities of smaller firms were not monitored, and these alliances are mostly the local ones.

The third aspect could be the influence of the earth's topography.
As shown in Figure~\ref{world}, the location of firms with \rud activity is highly restricted to only a few global hot spots.
Thus, the distribution of distances could be skewed because of the topography between these hot spots.
In particular, we assumed that the size of the Atlantic Ocean should have a notable influence since we observe a strong presence of edges between firms located at the US East Coast and firms located in the EU.
To test for this possible influence, we adjusted the observed distances using the probability of finding an edge of a certain distance by the calculated \knd values.
For each estimated bin $b$, the normalization factor $N_b$ is the fraction of the \knd value of the observed network and the density of the $50^{th}$ percentile of the random simulations: 
\begin{equation}
N_b = \frac{\hat{K}_n(d)_b}{\hat{K}_n(d)_b^{50^{th}}}.
\end{equation}
Thus, $N_b$ shows how strongly the observed \knd value deviates from a median outcome.
For each bin, the number of entries are then adjusted to the calculated $N_b$. 
As an example, consider that the first bin has 382 entries and its $N_b$ value is 3.55. 
We randomly pick an entry and append it to the existing ones until there are $382\cdot3.55=13561$ entries. 
For bins with $N_b <1$, entries are randomly deleted until their number matches the normalized number.
However, also this adjustment did not change the outcome either:
As shown in Figure~\ref{pdf-plots}, all different distributions (adjusted and not) deviate from what would be expected by a power law, but remain broad with heavy tails. 
\begin{figure} 
\centering
\subfigure[Total network with confidence bands]{
\includegraphics[width=.47\textwidth]{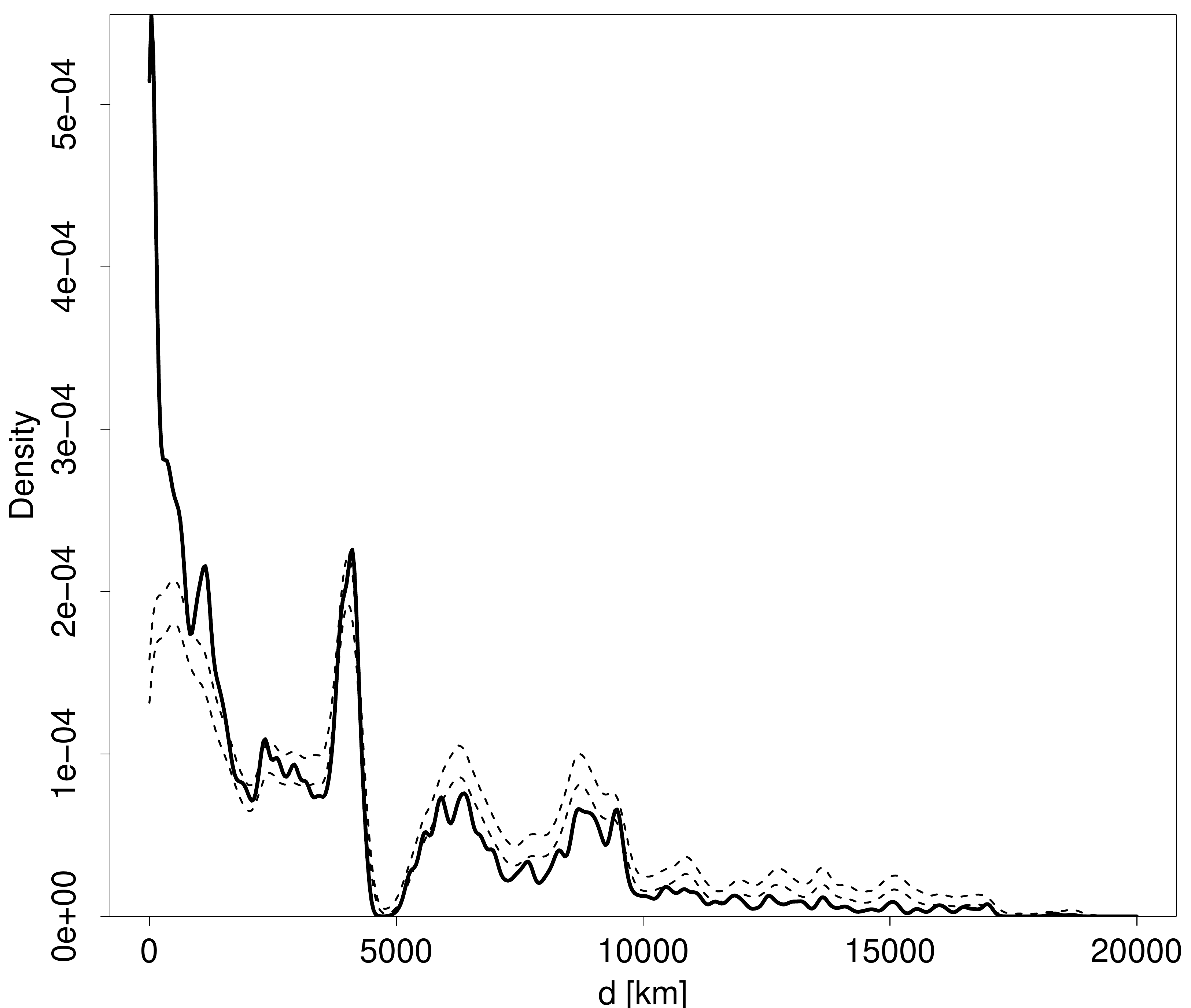}
\label{kde-do}
}
\hspace{0cm}
\subfigure[Time periods]{
\includegraphics[width=.47\textwidth]{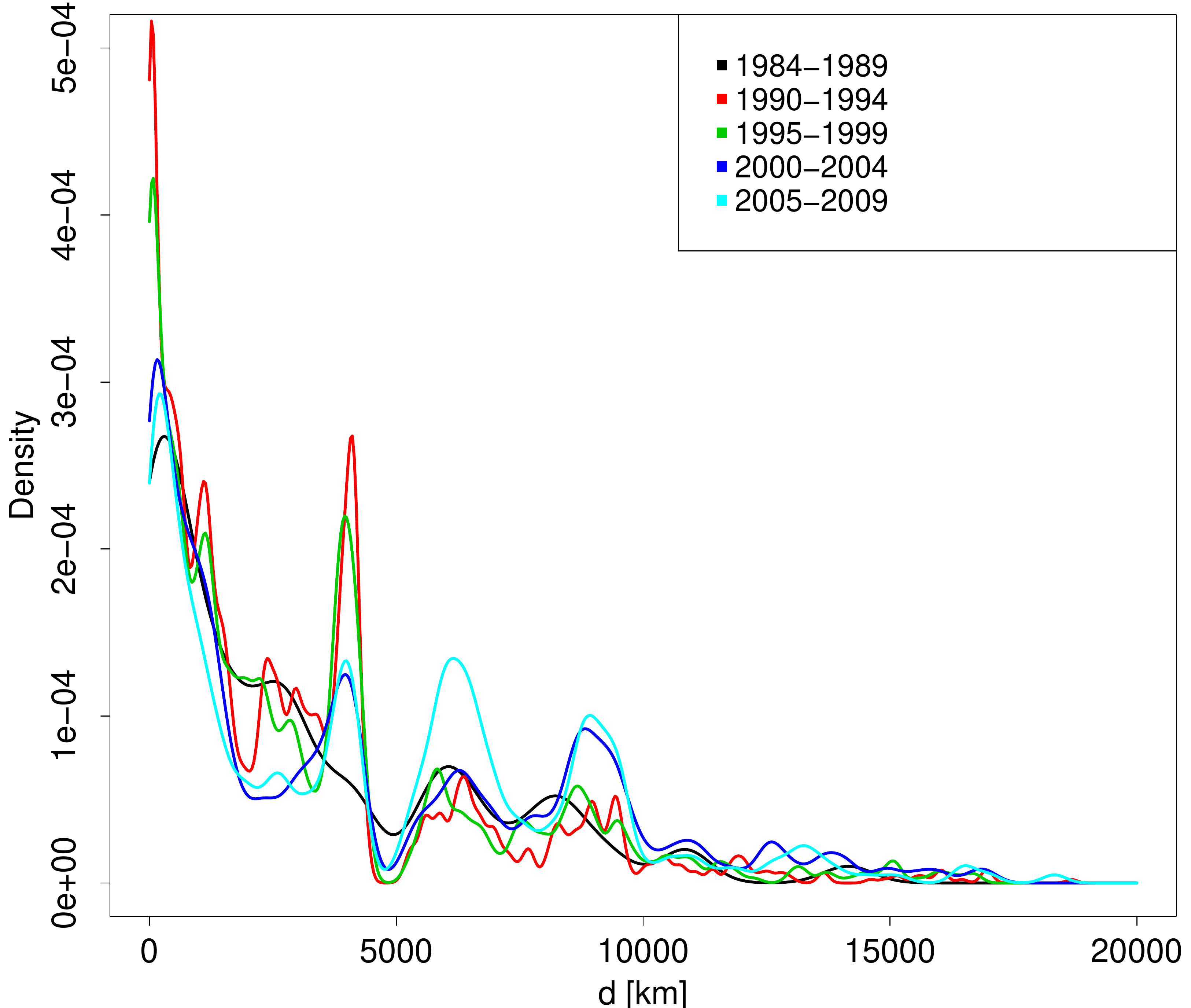}
}
\subfigure[SIC industries]{
\includegraphics[width=.47\textwidth]{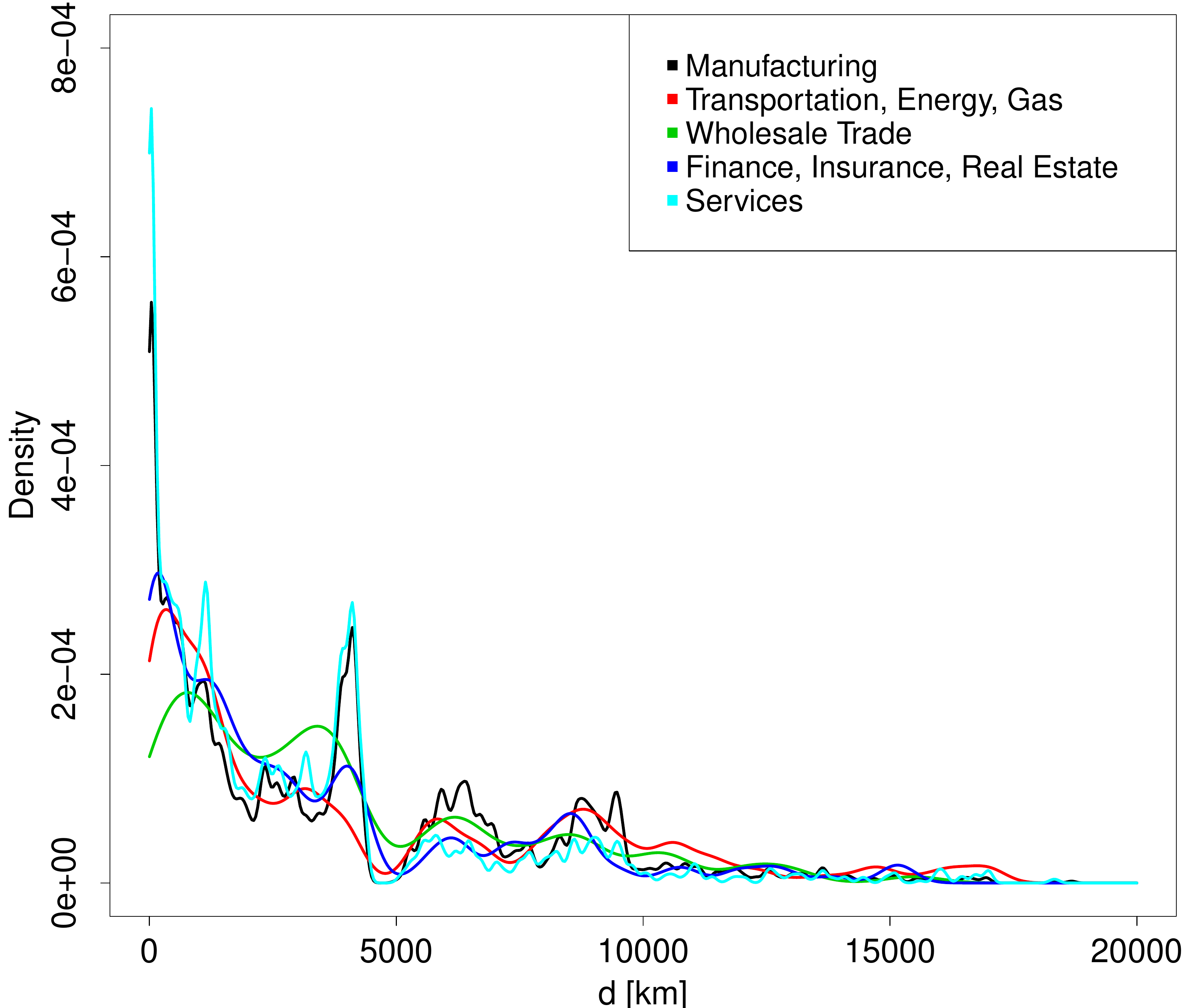} 
}
\subfigure[Firm size]{
\includegraphics[width=.47\textwidth]{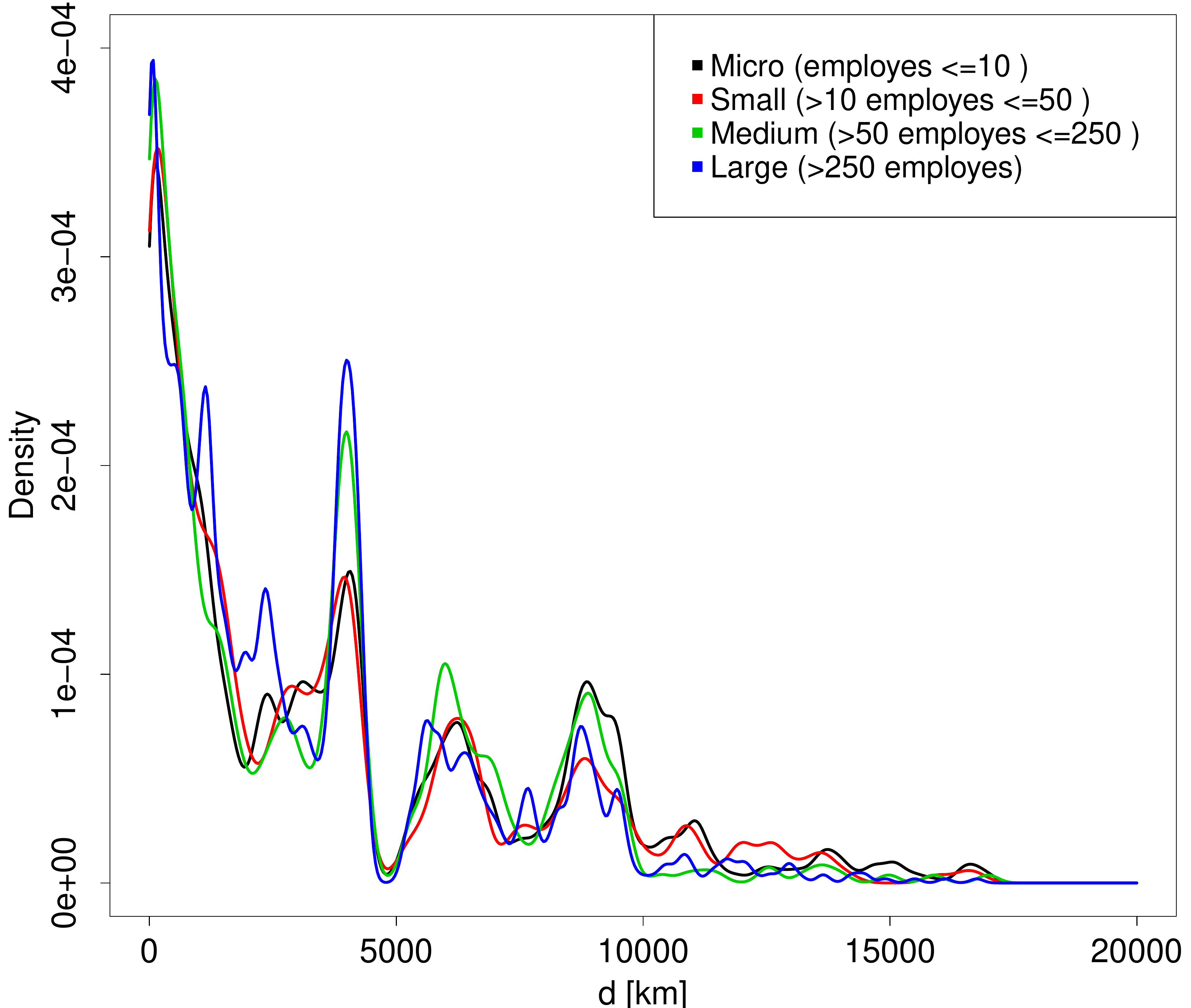} 
}
\caption{$\hat{K}_n(d)$ values for different \rud networks. Networks in sub-figure c are only plotted for number of edges E>100. } 
\label{kde}
\end{figure} 

Of course, failing to observe a power law for the aggregated distances does not mean that the distance distribution does not contain any information. 
In Figure~\ref{kde} we plot the $\hat{K}_n(d)$ values of the observed distances for a) the full \rud network, b) its decomposition into different time periods, c) its decomposition into industrial sectors, and d) its decomposition into firm sizes. 
In all these plots, we observe a peak at the 4000 km level which stands for the strong East Coast–West Coast collaboration network.
The peak at 6,000 km represents East Coast-EU collaborations while the peak at the distance between 7,000 and 9,000 stands for collaborations between EU- and East Coast firms and to firms located in Asia. 

More interesting, however, is the peak around the 0 km interval. 
Almost 6 \% of the collaborations in the full network take place at a very narrow scale below 20 km, and 12 \% are located at a distance between 0 and 100 km. 
This shows, that despite globalization and communication technologies, the direct surrounding area is a very important place for firms to find \rud collaborators. 

The confidence bands in Figure~\ref{kde}~(a) show that these peaks deviate from the random case.
In particular, the probability of finding collaborations at a short distance is much higher than expected from a random network. 
For the distance between zero and 1300 km, the $\hat{K}_n(d)$ values of the observed network exceed the upper confidence band.
The distances of the East Coast-West Coast collaboration network lie within the confidence bands while we observe less collaborations at larger distances than expected from a random network. 
The observation of a high probability for local and East Coast-West Coast collaborations is very robust. 

As discussed above, literature suggests a shift towards international \rud collaborations as the effect of globalization. 
This shift should be observed in our analysis, given that we can analyze changes in the collaboration activity between 1984 and 2009.  
By studying the network separately during consecutive time periods (4 year intervals between 1984 and 2009), 
a clear tendency towards larger distances in later time periods 
can indeed be observed in Figure~\ref{kde}~(b).
The rising of Asia displays in a higher peak at the 10,000 km level, at the expense of short distance and East Coast-West Coast collaborations.
Furthermore, the last time period from 2005 to 2009 
 is characterized by a strong focus on transatlantic collaborations. 
While it might be assumed that this is the outcome of pre-crisis collaborations of the finance industry, actually more than 70 \% of these realizations origin from manufacturing firms. 
The first  time period, 1984 - 1989, deviates form the described tendency towards larger distances in later time periods as it shows the lowest $\hat{K}_n(d)$ values at short distances of all observed five periods and does not show a peak at the 4,000 km level.
However, the first time period contains data for only 79 collaborations, so this result should be interpreted carefully.
In order to test for differences between the $\hat{K}_n(d)$ values of the five periods and the total network we applied the two sided Kolmogorov-Smirnov tests (KS-test).
Except for the time periods of 1995-1999 and 1984-1989, all the pairwise two sided KS-tests are not significant at a 5 \% significance level, indicating that the distribution of distances for these time periods differs from the distribution of the total network. 
\begin{figure} [t]
\begin{center}
\includegraphics[width=.45\textwidth]{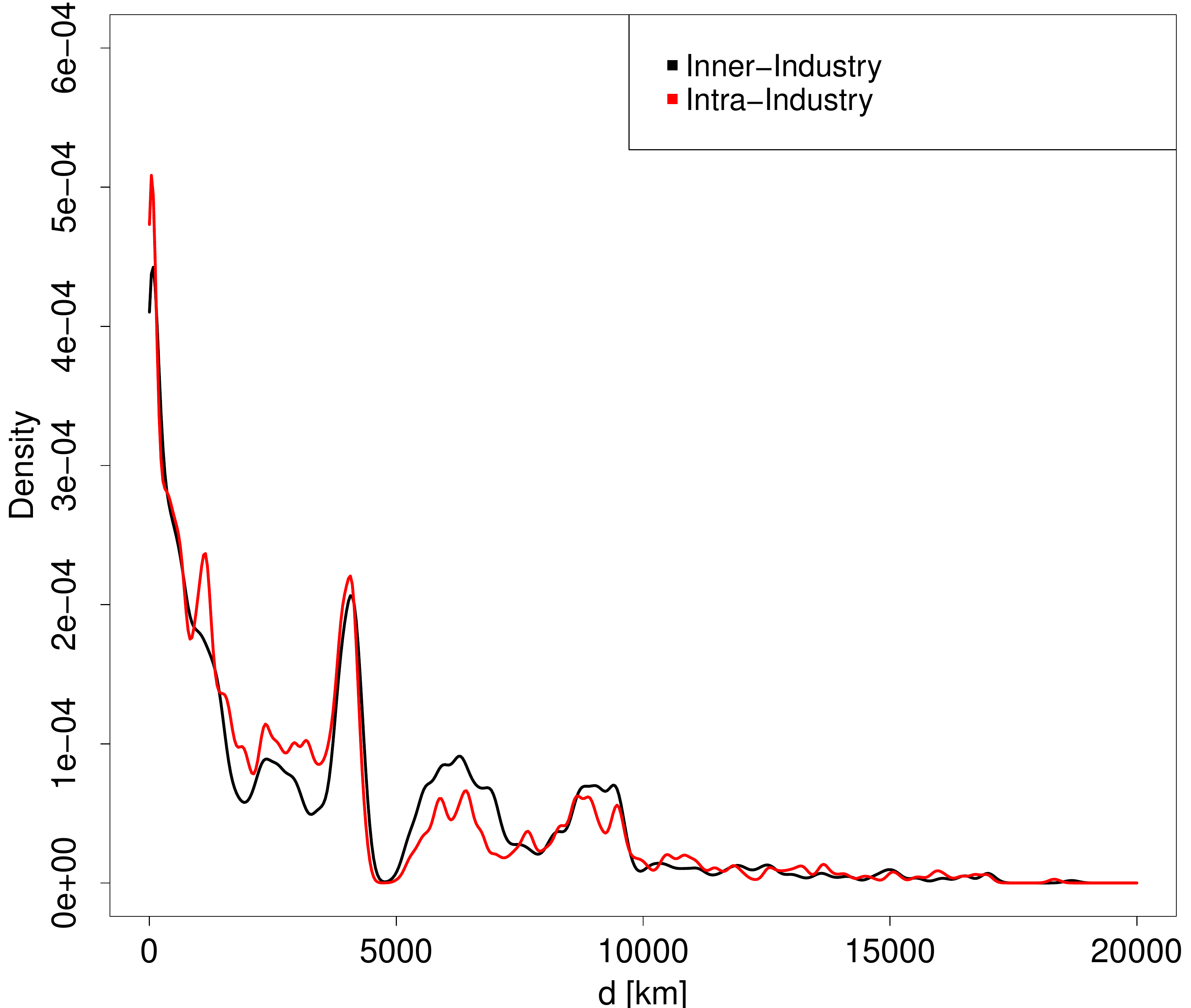}
\end{center}
\label{density_sic_same}
\caption{$\hat{K}_n(d)$ values for inner-industry and intra-industry collaborations.} 
\end{figure}

On the firm specific level, the scope of business matters more than the firm size.
Figure~\ref{kde}~(c) shows that $\hat{K}_n(d)$ values differ between different SIC codes.
On the one hand, collaborations of 
the two SIC industries Transportation, Energy, Gas, and Wholesale, Trade are relatively equally distributed in space. 
This results from the fact that the location choice of companies within the first industry is often spatially bounded to the location of natural resources, while the location of firms in the latter industry follows the location of urban agglomeration.
On the other hand, collaborations of service firms show the highest probability for short distances. 
This highlights the importance of spatial proximity in \rud activities of service firms~\cite[]{Muller200964}. 
The two sided KS-test is not significant for all investigated SIC codes except for the Finance, Insurance and Real Estate industry (5 \% significance level).

Since it is plausible to assume that the detected deviations between the industries result from their different tendency towards intra-industry cooperation, we calculated the  $\hat{K}_n(d)$ values separately for cooperation where the two partners have the same SIC code and a different SIC code, respectively. 
But, the differences between intra- and inner-industry collaborations are less pronounced than expected. 
The KS-tests detect a deviation from the total network only for inner-industry collaborations. 
Testing for deviations from the total network using the Mann-Whitney test is negative for both cases (5 \% significance level).

With respect to the firm size, we do not observe a distinctive difference for short distance collaborations in comparison to the other time periods and SIC codes.
Only at the 4000 km level, large and medium sized firms show a higher probability in comparison to micro and small sited firms. 
Concerning the KS-test, only micro and large sized significantly deviate from the total network (5 \% significance level). 

To summarize, all of the aforementioned results highlight the ambiguity of distance in economic cooperation.
The distribution of distances is multi-modal and differs between different industries and between different time periods, but it is comparatively similar for different classes of firm sizes. 
\begin{figure} [t]
\includegraphics[width=.48\textwidth]{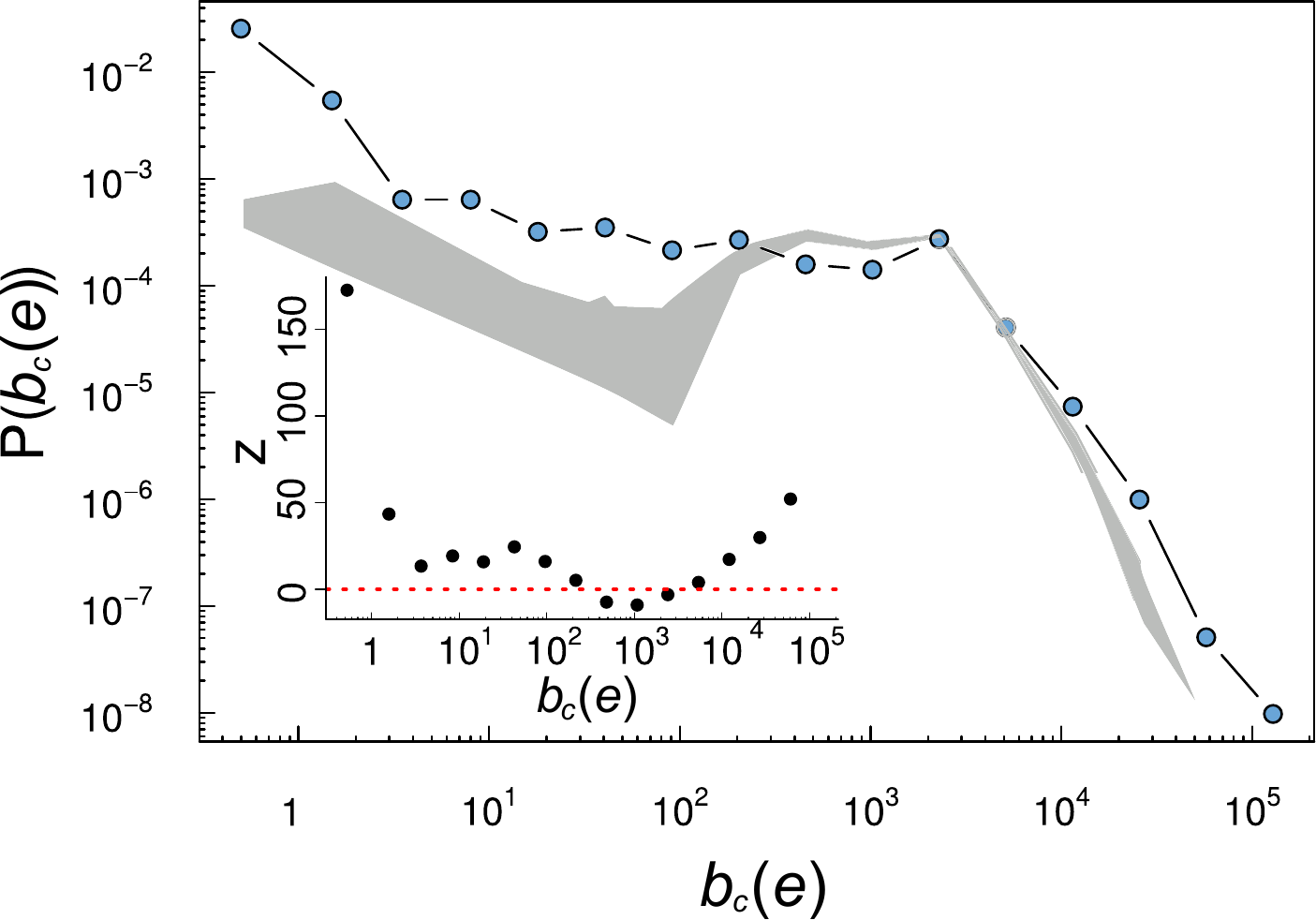}
\hfill
\includegraphics[width=.48\textwidth]{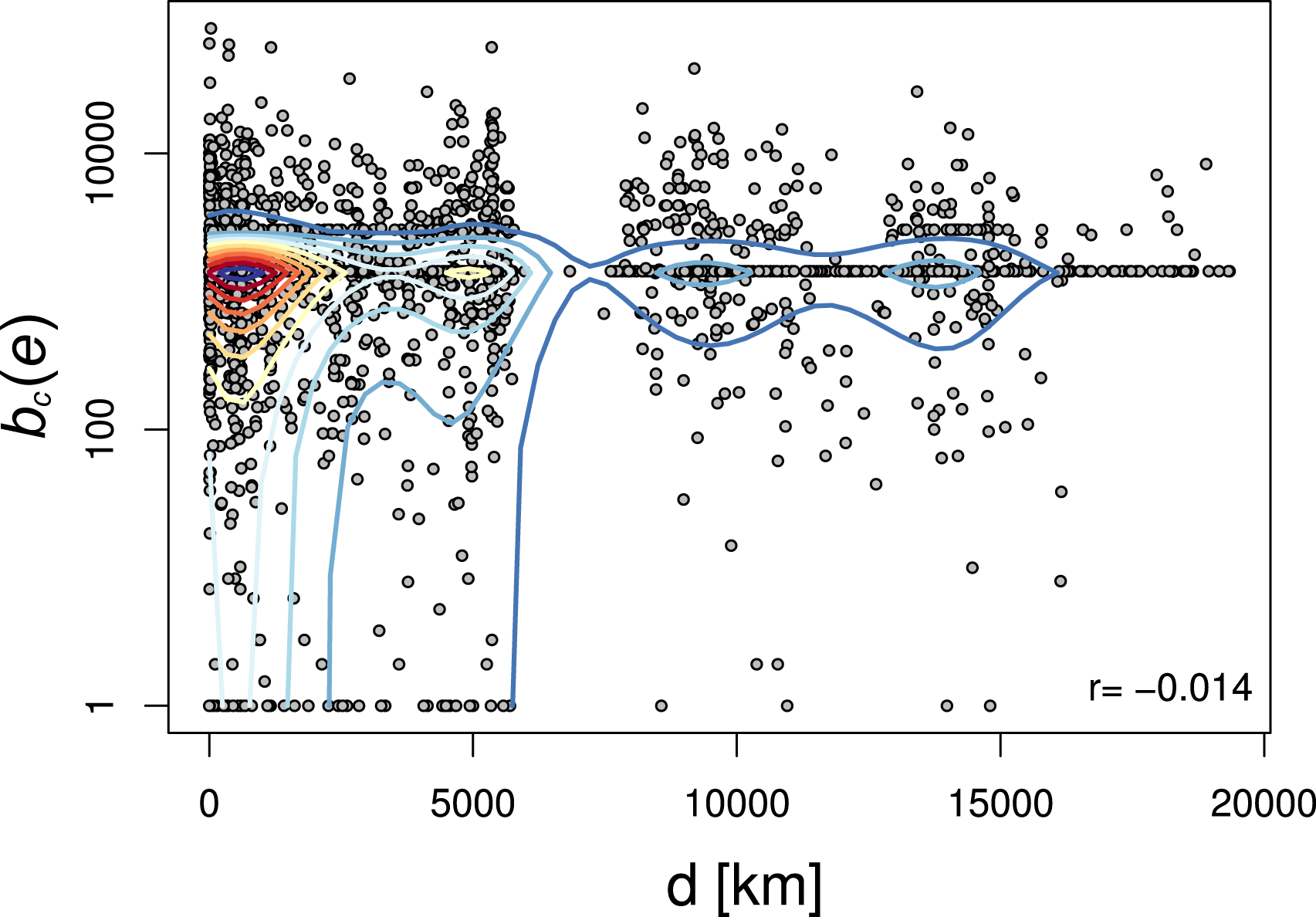}
\caption{Global pipeline properties. Left: Distribution of the $b_c(e)$  values for the empirical \rud
network (dots) and for the ensemble of randomly generated networks (shaded area). Inset: z-score
of the deviation between these two distributions. Right: Scatter plot of the $b_c(e)$  values of all edges
in the largest connected component of the \rud network versus their length. The contour lines
highlight regions with high concentration of points.
} 
\label{ebc}
\end{figure}

In what follows, we are going to test the applicability  of the \emph{local buzz and global pipelines} hypothesis to the observed \rud network.
Despite the absence of a power-law distribution of distances, it is possible that the \rud network follows   a  \emph{local buzz and global pipelines} structure, but with more emphasis on global pipelines. 
As discussed in section \ref{sec:role}, \emph{local buzz and global pipelines} properties can be measured by means of three network characteristics:
Significant concentration of collaborations at short distances, heavy-tailed distribution of $b_c(e)$  values and correlations between distances and $b_c(e)$  values.

Starting with the first characteristic that represents the \emph{local buzz} properties, the discussed outcome of the \knd plot in Figure~\ref{kde-do} has shown that the observed \rud network clearly fulfills this specification: 
Edges with short lengths have a significant higher probability than expected from a random outcome, while edges with larger lengths  do not exceed the confidence bands. 
All these are consistent with the existence of \emph{local buzz}.

But what about the pipelines? Do they exist, and if so how global are they?
As shown in Figure~\ref{ebc}, the distribution of $b_c(e)$  values is, indeed, heavy tailed.
In order to assess its significance, we compare it to the $b_c(e)$  distribution we would obtain from a set of random networks.
We repeat this step 1000 times, and for every bin $i$ of the distribution we calculate the z-score defined as ${\rm z}_i=({\rm b_c(e)}_i - \left<{\rm b_c(e)}_i^{*}\right>)/\sigma_{{\rm b_c(e)}_i^{*}}$, where $\left<{\rm  b_c(e)}^*\right>$ is the average value of the  $b_c(e)$ for the ensemble of random networks, and $\sigma_{{\rm  b_c(e)}_i^{*}}$ the standard deviation.
The z-score plot shows high positive deviations for both very small and very large  $b_c(e)$ values. 
The deviation towards large values indicates that the tails are, indeed, ``heavier'' than expected by a random topology.
In sum, this indicates that the \rud network contains central edges that act as pipelines, bridging separated clusters.
But, this alone does not mean that global pipelines do exist, since it contains no information of how ``global'' the edges with high  $b_c(e)$ values are.

While the observed \rud network  fulfills the first two properties of the \emph{local buzz and global pipelines} model, we do not find any significant correlation between the $b_c(e)$ values and the length of edges. 
The large deviation for small  $b_c(e)$ values points towards  areas with many more local edges with high  $b_c(e)$ values than what would be predicted by a random topology. These could just be isolated components in the global network, and their presence may only create some unwanted bias in the overall statistics.

Therefore, in order to correct for this, from the empirical \rud network we extract the largest connected component, and we study whether any correlation exists between the  $b_c(e)$ values and the length of the edges.
However, even after this correction the overall pattern remained the same, while the Pearson correlation coefficient between these two link properties is just $r=-0.014$, and as can be even visually seen in Figure~\ref{ebc} there is no traceable correlation.
So, we conclude that the global \rud network is \emph{not} characterized by \emph{local buzz and global pipelines} \emph{but} by \emph{local buzz and gloCal pipelines}: 
Both local buzz and pipelines exist, but the pipelines do not only bridge clusters at a global scale, they are \emph{simultaneously important at all levels}  or, in other words, at a gloCal level.

However, similar to our discussion about the distribution of distances, this finding might be influenced by our dataset which is biased towards \rud activities of large US/EU firms. 
In addition, because we study \rud collaborations using a quantitative network approach, and in our analysis we do not consider the \emph{value} of a collaboration for the firms involved, we cannot exclude that  collaborations at larger distances are more important for knowledge creation and exchange than the local ones.
But, this is something we cannot test with our data and further quantitative and qualitative  work is necessary to check for this possible influence.

\section{Conclusions}
This paper studies the spatial component of global \rud collaborations by means of a complex network approach.
Our empirical results provide evidence for the ambiguity of distance in economic cooperation which is also suggested by the existing literature.
By pushing the geographic scale of our analysis to a micro geographic level, we are able to observe this ambiguity at all spatial scales, simultaneously. 
This approach allows for new insights into the spatial component of global \rud collaborations and leads to both well described and surprising outcomes.

One of the expected outcomes is the strong presence of collaborations at short distances.
12 \% of all \rud collaborations take place  at a distance between 0 and 100 km and 6 \% are even located at a very narrow scale below 20 km. 
Our results support the idea that spatial proximity favors \rud collaborations between firms as it eases the buildup of trust and reduces the costs for communication. 
The high importance of short-distance collaboration is observable for all decompositions of the \rud network into different time periods, industrial sectors and  classes of firm sizes. 
By means of these decompositions, we can also observe other expected outcomes: 
As suggested by literature, \rud collaborations of service firms are, in comparison to other industries, even more concentrated at short distances.  
Furthermore, the \rud network shows a tendency towards larger distances over time which can be seen as an effect of globalization.

On the other hand, we also find results that deviate from what is reported in literature.
First, we do not find any evidence for a power-law distribution of distances in \rud collaborations.
While such a  distribution has been reported for many other similar networks (e.g. scientific collaboration, human mobility, ownership-networks), \rud collaborations at large distances are much more frequent than expected from  a power-law distribution.
By contrast, even after normalizing the possible influence of the earth's topography, the regional level is less pronounced in comparison to similar networks. 
Comparing the share of international collaboration, we find out that notwithstanding the high importance of the local level, \rud collaboration is a more international orientated process in comparison to scientific collaboration. 

The observed co-existence of local and global collaborations has been described theoretically in the model of \emph{local buzz and global pipelines} by \cite{Bathelt}.
The present paper has discussed how this model can be tested quantitatively by means of three network characteristics.
The studied \rud network shows both a significance of collaborations at short distances and a heavy-tailed distribution of their edge betweenness centrality values.
This supports the idea of the model of Bathelt et al. that local routines lead to a higher collaboration at the local level and that some firms are important for building bridges of collaboration between clusters of  firms.
However, we do not observe any correlation between the geographical length and the betweenness centrality of an edge.
Although the regional level is less pronounced in the distribution of distances according to what would be expected by a power-law distribution, it is equally important for connecting local clusters of firms.
Thus, from the observed statistical results  we conclude that the \rud network is characterized by  \emph{local buzz and gloCal pipelines}.
 
In sum, the spatial component of \rud networks seems to be more heterogeneous than reported for similar networks of collaboration.
While the existing literature provides different theories that support parts of our findings other results seem rather puzzling.    
Thus, we foresee a large demand for  both observing further \rud networks form a micro-geographic perspective  and for building theories that can explain the observed outcomes.

\section{Acknowledgements}
AG and FS acknowledges support by the EU-FET project MULTIPLEX 317532.  TS acknowledges support by the German Academic Exchange Service (DAAD). Early work on this subject was supported by SNF through grant 100014\_126865

\bibliographystyle{authordate2} 
\bibliography{spatial-rnd}{}
\end{document}